\documentclass[twocolumn]{aa}
\usepackage{graphicx}
\usepackage{natbib}

\usepackage{txfonts}
\begin{document}
\title{Astrometric and photometric monitoring of GQ Lup and
its sub-stellar companion\thanks{Based on observations collected
at the European Southern Observatory, Chile in runs 073.C-0164, 
075.C-0710C, 076.C-0339B, 077.C-0264B, 078.C-0552B, and 078.C-0535A.}}

\author{Ralph Neuh\"auser\inst{1} \and
Markus Mugrauer\inst{1} \and
Andreas Seifahrt\inst{1} \and
Tobias O.\,B. Schmidt\inst{1} \and
Nikolaus Vogt\inst{2,3} }

\institute{
Astrophysikalisches Institut, Universit\"at Jena, Schillerg\"asschen 2-3, 07745 Jena, Germany
\and
Departamento de F\'isica y Astronom\'ia, Universidad de Valpara\'iso,
Avenida Gran Breta\~na 1111, Valpara\'iso, Chile
\and
Instituto de Astronom\'ia, Universidad Catolica del Norte, Avda. Angamos 0610,
Antofagasta, Chile
}

\offprints{Ralph Neuh\"auser, \email{rne@astro.uni-jena.de}}

   \date{Received 16 Aug 2007; accepted 4 Dec 2007 } 

  \abstract
{
Neuh\"auser et al. (2005) presented direct imaging evidence for a 
sub-stellar companion to the young T~Tauri star GQ Lup.
Common proper motion was highly significant, 
but no orbital motion was detected.
Faint luminosity, low gravity, and a late-M/early-L spectral type indicated 
that the companion is either a planet or a brown dwarf.
}
{
We have monitored GQ Lup and its companion in order to
detect orbital and parallactic motion
and variability in its brightness.
We also search for closer and fainter companions.
}
{
We have taken six more images with the VLT Adaptive Optics instrument NACO
from May 2005 to Feb 2007, always with the same calibration binary from Hipparcos
for both astrometric and photometric calibration.
By adding up all the images taken so far, we search for
additional companions.
}
{
The position of GQ Lup A and its companion compared to a nearby
non-moving background object varies as expected for parallactic 
motion by about one pixel ($2 \cdot \pi$ with parallax $\pi$).
We could not find evidence for variability of the GQ Lup companion
in the K$_{\rm s}$-band (standard deviation being $\pm 0.08$ mag), 
which may be due to large error bars.
No additional companions are found with deep imaging.
}
{
There is now exceedingly high significance for common proper
motion of GQ Lup A and its companion. In addition, we see for the
first time an indication for orbital motion ($\sim 2$ to 3 mas/yr decrease
in separation, but no significant change in the position angle),
consistent with a near edge-on or highly eccentric orbit. 
We measured the parallax for GQ Lup A to be $\pi = 6.4 \pm 1.9$ mas 
(i.e. $156 \pm 50$ pc) and for the GQ Lup companion to be
$7.2 \pm 2.1$ mas (i.e. $139 \pm 45$ pc), both consistent with
being in the Lupus I cloud and bound to each other.
}

\titlerunning{Astrometric and photometric monitoring of the GQ Lup system}


\maketitle

\section{Introduction: GQ Lup A and its companion}

Based on three epochs of imaging data spanning five years,
Neuh\"auser et al. (2005, henceforth N05) presented evidence
that the $\le $ few Myr young T Tauri star GQ Lup 
has a co-moving companion with K$_{\rm s} \simeq 13.1$ mag
about 0.7$^{\prime \prime}$ west ($\sim 100$ AU at $\sim 140$ pc) of the primary star. 
A low-resolution spectrum gave evidence that it has a late-M to early-L spectral type. 
Temperature and luminosity can be used to estimate the mass via
theoretical evolutionary models. They are, however, very uncertain at
young ages (up to at least $\sim 10$ Myr; Chabrier et al. 2005).
Using the Wuchterl \& Tscharnuter (2003) model extended to 
planetary masses and few Myr of age, the GQ Lup companion could
be a 1 to 3 M$_{\rm Jup}$ object;
whereas for the Tucson group models (Burrows et al. 1997), 
it is few to $\sim 30$ M$_{\rm Jup}$;
and for the Lyon group models (Baraffe et al. 2002), 
it is few to $\sim 40$ M$_{\rm Jup}$.
A comparison with the GAIA-1 model atmospheres indicated low gravity
and, hence, a young age and very low mass (N05).
Marois et al. (2007) re-analyzed archival HST and Subaru data to study the
spectral energy distribution of GQ Lup A and its companion, showing possible
evidence for an excess in the L- and R-bands (for the companion), possibly 
due to a disk and H$\alpha$ emission; a fit of the data to 
the dusty GAIA model atmospheres confirmed the temperature and luminosity of
the companion given in N05 and revised its radius to $0.38 \pm 0.05$~R$_{\odot}$,
yielding a mass between 10 and 20 M$_{\rm Jup}$.
McElwain et al. (2007) then obtained higher-resolution spectra (R $\simeq 2000$)
of the GQ Lup companion in the J- and H-bands with the integral field spectrograph OSIRIS 
at Keck and found a slightly higher temperature (M6-L0) than in N05 (M9-L4),
explained by the fact that H$_{2}$ collision-induced absorption is important
for low gravity objects according to Kirkpatrick et al. (2006), but not 
considered in N05.

Higher-resolution spectra in the J-, H-, and K-bands
taken with VLT/Sinfoni compared to GAIA-2
models could better constrain the parameters of the companion:
the temperature is $2650 \pm 100$ K,
the gravity $\log~g = 3.7 \pm 0.5$ dex, 
and the 
radius $3.5^{+1.50} _{-1.03}$ 
R$_{\rm Jup}$ (Seifahrt et al. 2007).
Hence, its mass can be as low as few M$_{\rm Jup}$, but also much higher.
Comparing its parameters with 2M0535 A \& B, where masses have been
determined dynamically (an eclipsing double-lined binary brown dwarf
in Orion, similar age as GQ Lup, Stassun et al. 2006), could give an
upper mass limit of 35 M$_{\rm Jup}$ (Seifahrt et al. 2007).
Thus, the companion to GQ Lup can be regarded as a {\em planet
candidate} according to the best guess value of its mass, 
which is at or below the brown dwarf desert ($\sim 30$ M$_{\rm Jup}$;
Grether \& Lineweaver 2006), proposed as the deviding line between planets 
and brown dwarfs. Also, it is possible that the true 
mass is below 13 M$_{\rm Jup}$, a more conservative upper mass limit
for planets.


The large error in the luminosity of the companion, 
which is used for the mass estimate from evolutionary models, 
is mostly due to a large distance uncertainty,
assuming that the object is in Lupus I 
($\pm 50$ pc, N05). 
Hence, a direct parallax measurement would yield a stronger
contraint on luminosity and, hence, mass.

To finally confirm that the fainter object near GQ Lup is really
a bound companion (rather than, e.g., another member of the Lupus I cloud
not orbiting GQ Lup), one would need to see orbital motion.

For both measurements, parallax and orbital motion, we have
monitored GQ Lup and its companion from 2005 to 2007 by taking
six new images with Adaptive Optics. In Sects. 2 \& 3, we explain
the observations, data reduction and astrometric results. 

%

We also use the data to monitor the brightness of GQ Lup A and
its companion and search for photometric variability (see Sect. 4).
We note that Seifahrt et al. (2007) found emission lines in the
near-infrared spectra of the companion, indicative of accretion,
so some variability is expected.
We then add up all imaging data thus far available to obtain a very
deep and high-dynamic-range image to search for additional, fainter
and/or closer companions (see Sect. 5). 
We note that both Debes \& Sigurdsson (2006)
and Boss (2006) argued that if the GQ Lup companion is a planet,
it should have been moved to its current large separation ($\sim 100$ AU)
from an originally closer orbit by an encounter with another more massive
proto-planet, which may be detectable.
We use the newly determined parameters to re-estimate some physical
parameters of the GQ Lup companion in Sect. 6 and summarize our results
at the end.

\section{Observations with VLT / NACO}

We observed GQ Lup A and its companion six times from May 2005
until Feb 2007 in order to detect and monitor 
possible changes in separation, position angle, and
brightness. See Table 1 for the observations log,
where we also include the first VLT imaging
observation from June 2004, which we re-reduced here.

All observations were done with the European Southern Observatory (ESO)
Very Large Telescope (VLT) instrument Naos-Conica (NACO; Rousset et al. 2003) in the K$_{\rm s}$-band
around 2 $\mu$m: 200 images (NDIT) with 0.347 seconds (DIT) each, 
the shortest possible integration time,
all added up automatically to one frame (without moving the telescope).
We then took 15 frames (NINT) with GQ Lup on slightly different positions 
(jitter or dither technique), but 18 frames in May 2005 and 27 in June 2004.
We always used the S13 camera (pixel scale roughly 13 mas/pixel)
and the double-correlated read-out mode.
For the astrometric calibration binary HIP 73357, observed within one hour\footnote{In the
Feb 2007 service mode observation, HIP 73357 B was inside the field-of-view only
in two out of five frames, so we could not use it properly. Instead, we 
used HIP 73357 observations done with the identical set-up a few nights later on
1 March 2007 (run 078.C-0535A), where 43 useful high-S/N frames were taken 
($80 \times 0.347$ seconds each) by some of us. We used them to calibrate the
19 Feb 2007 observations of GQ Lup, because there was no intervention
into NACO in the meantime, as confirmed with ESO.}
of the GQ Lup observations with the same set-up, we took five frames,
i.e., five jitter positions with $200 \times 0.347$ seconds each
(but six such images in May 2005 and four each in May and July 2006).
For the raw data reduction, we subtracted a mean dark from all the science frames 
and flats, then devided by the mean normalized dark-subtracted flat and subtracted 
the mean background.

Compared to the data reduction for N05, we now use a new, improved IDL routine for
the subtraction of the point spread function (PSF) of the bright GQ Lup A, 
before measuring the companion. The new routine centers the bright star PSF with
sub-pixel precision:
after rotating the PSF of GQ Lup A around its center 179 times in steps of 2 degrees, 
we subtract the rotated image from the original image,
and then take the median of all resulting images. This gives the final PSF-subtracted
image, which we use to measure the companion (for astrometry and photometry).

We re-reduced the GQ Lup data from June 2004 (published in N05) and thereby
obtained slightly better (larger) values for separation and brightness,
consistent with N05 within the errors. The new values are given in Table 2.
The new values for GQ Lup for 2004 use the same calibration results as in N05
for pixel scale and detector orientation (Table 1), since it was not necessary
to re-reduce the 2004 calibration binary observation.

\section{Astrometry}

Since GQ Lup has right ascension 15h 49m, we always placed one observation
in May (2005 and 2006), when it is visible the whole night,
two observations three months earlier (Feb 2006 and 2007) and
two more observations three months later (July/Aug 2005 and 2006), 
when GQ Lup is still visible at the end or start of the night, respectively,
so that we can use the data not only for proper motion and photometric monitoring, 
but also for trying to determine the absolute or relative parallaxes of GQ Lup A and its companion.

For the astrometric calibration, we observed 
the Hipparcos binary star HIP 73357 in each run. 
See the online appendix for more information
and Table 1 for the results.

\begin{table*} 
\begin{tabular}{llccc|rr|rr}
\multicolumn{9}{c}{\bf Table 1. VLT/NACO observation log and astrometric calibration results from HIP 73357.} \\ \hline
Epoch & Date of     & No. (a) & FWHM  & Re- & \multicolumn{2}{c}{pixel scale in [mas/pixel]}  & \multicolumn{2}{c}{detector orientation in [$^{\circ}$]} \\
year  & observation & images  & [mas] & mark  & value \& abs. err. & change \& rel. err.  & value \& abs. err. & change \& rel. err. \\ \hline
2004.48 & 25 Jun 2004 & 27     & 64  & (b) & $13.23  \pm 0.05 $ & (b)                & $0.14 \pm 0.25$ & (b)    \\
2005.40 & 27 May 2005 & 18     & 74  &      & $13.240 \pm 0.050$ & $0     \pm 0.001$  & $0.21 \pm 0.33$ & $0     \pm 0.007$ \\
2005.60 & 7/8 Aug 2005 & 15    & 63  & (c) & $13.250 \pm 0.052$ & $0.010 \pm 0.002$  & $0.35 \pm 0.34$ & $0.146 \pm 0.014$ \\ 
2006.14 & 22 Feb 2006 & 15     & 66  & (c) & $13.238 \pm 0.053$ & $-0.002 \pm 0.003$ & $0.18 \pm 0.35$ & $-0.024 \pm 0.020$ \\ 
2006.38 & 20 May 2006 & 15     & 100 &      & $13.233 \pm 0.055$ & $-0.007 \pm 0.006$ & $0.39 \pm 0.36$ & $0.184 \pm 0.028$ \\
2006.54 & 16 Jul 2006 & 15     & 73  &      & $13.236 \pm 0.055$ & $-0.004 \pm 0.005$ & $0.43 \pm 0.36$ & $0.221 \pm 0.031$ \\ 
2007.13 & 19 Feb 2007 & 15     & 110 & (d) & $13.240 \pm 0.059$ & $0.000 \pm 0.010$  & $0.34 \pm 0.38$ & $0.128 \pm 0.044$ \\ \hline
\end{tabular}

Remarks: For each VLT/NACO observation, we list the pixel scale determined with its
absolute error and the change in pixel scale since the first new observation (2005.4),
always with the same astrometric standard star HIP 73357; same for detector
orientation in the last two columns. \\
(a) Each image consists of 200 exposures with 0.347 sec each,
for both GQ Lup and the astrometric calibration HIP 73357.
(b) Results from 2004.48 as in N05, no relative errors available, 
because another astrometric calibration binary was observed. 
(c) GQ Lup A in a non-linear regime (not saturated).
(d) GQ Lup A and HIP 73357 A in a non-linear regime (not saturated).
Calibration (for 19 Feb 2007 GQ Lup data) obtained on 1 Mar 2007 with HIP 73357,
see footnote 1.
\end{table*}

\subsection{Astrometric data reduction on GQ Lup}

One can combine all images taken within, e.g., one night by standard shift+add 
to obtain one image with very high S/N. We did this in N05, and have done the
same for the six new observations from 2005 to 2007. In the final image,
one can then determine the PSF photocenter of the bright star GQ Lup A and,
after PSF subtraction, also the PSF photocenter of the companion.
This results in measurements of separation in pixels and 
the position angle (PA) of the companion relative to star A.
The companion is always slightly north of west.
The error for the separation include errors from the Gaussian centering
on A and companion, as well as the error in the pixel scale determined;
the error for the PA includes centering errors and the error in the
north-south alignment of the detector during the observation.
In Table 2, we list the separation in pixels and arc sec
(computed with the pixel scales from Table 1) and also the
PA in degrees, as corrected with the detector orientation
given in Table 1 for each epoch.

\begin{table*}
\begin{tabular}{l|ccc|cc|cc|c}
\multicolumn{9}{c}{\bf Table 2. Separation and PA between GQ Lup and its companion.} \\ \hline
Epoch & separation & separation & sep. change  & Position Angle & Change in PA & \multicolumn{2}{c}{sign. against} & sign. for \\
year  & (absolute) & (absolute) & since 2005.4 & (absolute) & since 2005.4 & \multicolumn{2}{c}{being b/g (d)} & orbital \\
      & [pixel] (a)& [mas] (a)  & [mas] (b)    & [deg] (a)  & [deg] (b)    & re PA & re sep & motion (e) \\ \hline
2004.48 & $55.53 \pm 0.11$ & $734.7 \pm 3.1$ & (c)           & $275.48 \pm 0.25$ & (c) & 3.2 $\sigma$ & 4.5 $\sigma$ & -- \\
2005.40 & $55.52 \pm 0.13$ & $735.1 \pm 3.3$ & $0   \pm 0.5$ & $276.00 \pm 0.34$ & $0 \pm 0.042$ & -- & -- & -- \\
2005.60 & $55.34 \pm 0.20$ & $733.3 \pm 3.9$ & $-1.8 \pm 1.2$ & $275.87 \pm 0.37$ & $-0.13 \pm 0.14$ & 0.9 $\sigma$ & 1.1 $\sigma$ & 0.5 $\sigma$ \\
2006.14 & $55.13 \pm 0.12$ & $729.8 \pm 3.3$ & $-5.3 \pm 1.0$ & $276.14 \pm 0.35$ & $0.14 \pm 0.05$ & 2.4 $\sigma$ & 4.7 $\sigma$ & 4.2 $\sigma$ \\
2006.38 & $55.27 \pm 0.15$ & $731.4 \pm 3.5$ & $-3.7 \pm 1.0$ & $276.06 \pm 0.38$ & $0.06 \pm 0.12$ & 3.9 $\sigma$ & 5.2 $\sigma$ & 2.6 $\sigma$ \\
2006.54 & $55.40 \pm 0.30$ & $733.2 \pm 5.0$ & $-1.9 \pm 2.1$ & $276.26 \pm 0.68$ & $0.26 \pm 0.58$ & 4.1 $\sigma$ & 4.0 $\sigma$ & 0.0 $\sigma$ \\
2007.13 & $55.14 \pm 0.42$ & $730.0 \pm 6.4$ & $-5.1 \pm 4.6$ & $276.04 \pm 0.63$ & $0.04 \pm 0.50$ & 5.0 $\sigma$ & 3.1 $\sigma$ & 0.5 $\sigma$ \\ \hline
\end{tabular}

Remarks:
PA is from north over east to south, after correcting for the detector orientation. \\
(a) Absolute errors, (b) relative errors,
(c) no relative errors available, because another astrometric calibration binary was observed,
(d) significance in Gaussian $\sigma$ against background (b/g) hypothesis for each epoch 
compared to the first new epoch (2005.4) regarding (re) PA (Fig. 1) and separation (Fig. 2);
the background hypothesis is that the companion is a non-moving background object, 
in which case GQ Lup A would move away due to its known proper motion.
The significances compared from the three earlier observations from 1994 to 
2002 are 5.4, 7.0, \& 6.9 $\sigma$ for separation, respectively,
and 7.6 \& 8.3 $\sigma$ for PA of 2005.4 compared to 1999 and 1994, respectively.
(e) The significance for deviation of each separation value from the 2005.4 
value is given in the last column (see Fig. 2b).
It can be seen as evidence for orbital motion.
(d,e) Because the measurements are independent, we can add up the significances (properly)
and obtain altogether very high significance against the background hypothesis
and $\ge 5 \sigma$ for having detected orbital motion.
\end{table*}

To check the errors in separation and/or PA,
we not only measured the photocenters of GQ Lup and its companion
in the one full final high-S/N image after combining all 
images of one epoch (one night), but we also measured separation 
and PA in any of the 15 to 27 GQ Lup images, 
as well as in {\em bins} of 3, 5, 7, and 9 GQ Lup images.
With {\em bins} of, e.g., $x$ images, we mean here the combination of
$x$ individual images by shift+add, e.g., we can obtain three combined
images with five individual images added up (or e.g. 5 combined images
with 3 individual images added up). We can then determine separation
(and PA) in any of the combined images. The weighted mean of those values
is then the final measurement of separation (and PA) of that epoch,
and the standard deviation is the precision of that measurement.
We use the larger error values in Table 2 for the absolute errors.
Since the calibration binary HIP 73357 has changed its PA significantly
since 1991.25, the calibration errors in PA are relatively large, so that
they overwhelm real measurement errors. 
 
\begin{figure*}
\centering
\includegraphics[width=120mm,angle=270]{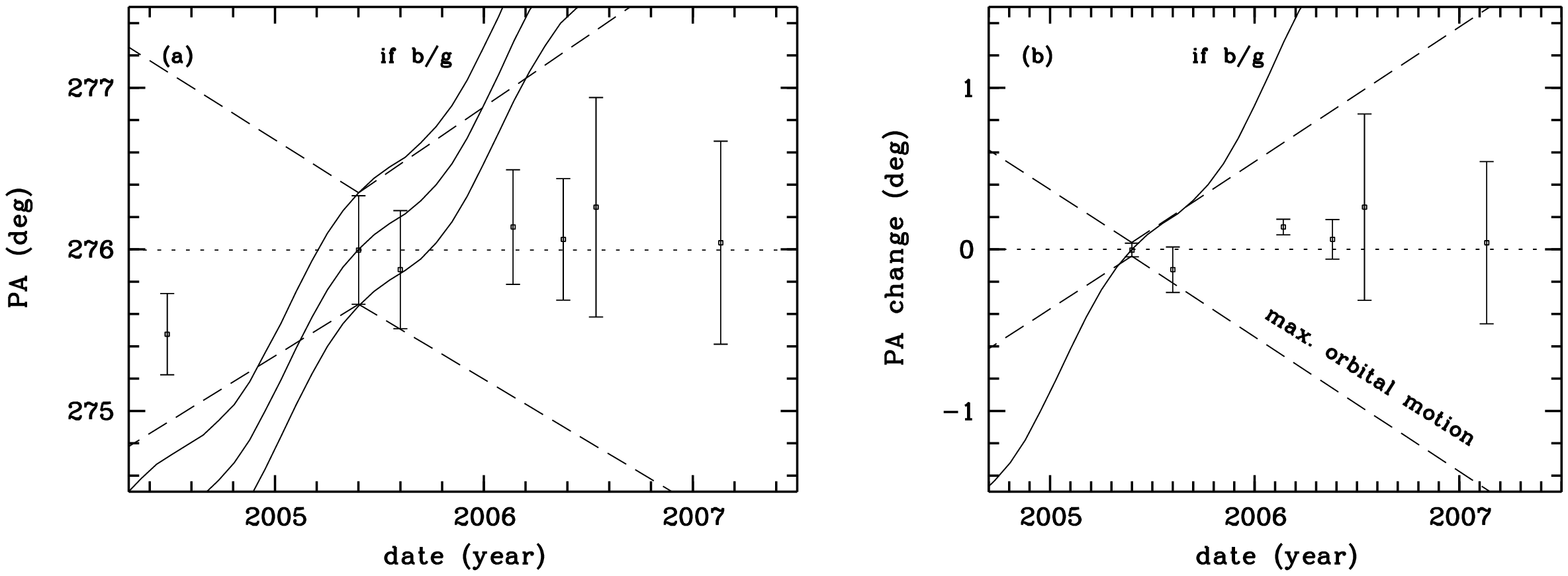}
\caption{PA of the companion relative to GQ Lup A versus time. 
(a) Left panel: absolute PA values with their absolute errors for all
NACO observations (Table 2).  
(b) Right panel: 
change in PA since May 2005 with relative errors, the error includes orbital motion
of the HIP 73357 calibration binary since 2005.4 only (not since 1991.25
as in the left panel). 
We also show the data point at the first new epoch 2005.4,
set to 0.0, with its error bar just from the Gaussian centering fit.  
For both (a) and (b), the straight dotted line is for constant PA,
as expected if both objects are bound.
To allow for orbital motion,
the maximum change in PA (if pole-on orbital plane) is indicated by
dashed lines. The data stay constant or vary slowly within those limits.
There is no strong evidence for orbital motion in PA;
the first data point from 2004 should have lower weight here, because it 
was obtained with different astrometric calibration binaries than in 2005 to 2007.
The full wobbled lines with strong positive slope 
indicate the change in PA, if the companion was background
($\pm$ expected errors from errors in proper motion and parallax
in panel a);
this hypothesis can be rejected with large significance, 
because the data points strongly deviate from this hypothesis.
}
\end{figure*}

We correct the separation measured in pixels and the PA measured for GQ Lup
and its companion with the calibration values given in Table 1.
In Table 2, we list the values for separation and PA between 
GQ Lup A and its companion.
In Figs. 1 \& 2, we plot the PA and separations versus time,
as observed and listed in Table 2. The precision of the
measurements decrease slightly from 2005 to 2006 and are
worst in 2007, where we have strong reflection effects and
a strong waffle structure (bad seeing, large FWHM). This decrease 
is at least partly due to the fact that the error in the astrometric
calibration binary from its possible orbital motion since 1991.25 
(Hipparcos) increases with time. The 2006 data for both GQ Lup and
HIP 73357 also have much lower S/N than those in 2004 and 2005.

For the last seven NACO measurements, the mean in separation between
GQ Lup A and its companion is $732.1 \pm 2.1$ mas and the
mean PA is $275.98 \pm 0.25 ^{\circ}$, i.e., slightly north of west.
This separation corresponds to $114 \pm 33$ AU 
at $156 \pm 50$ pc, the distance to GQ Lup A determined in Sect. 3.4.

\subsection{Proper motion and orbital motion}
 
\begin{figure*}
\centering
\includegraphics[width=120mm,angle=270]{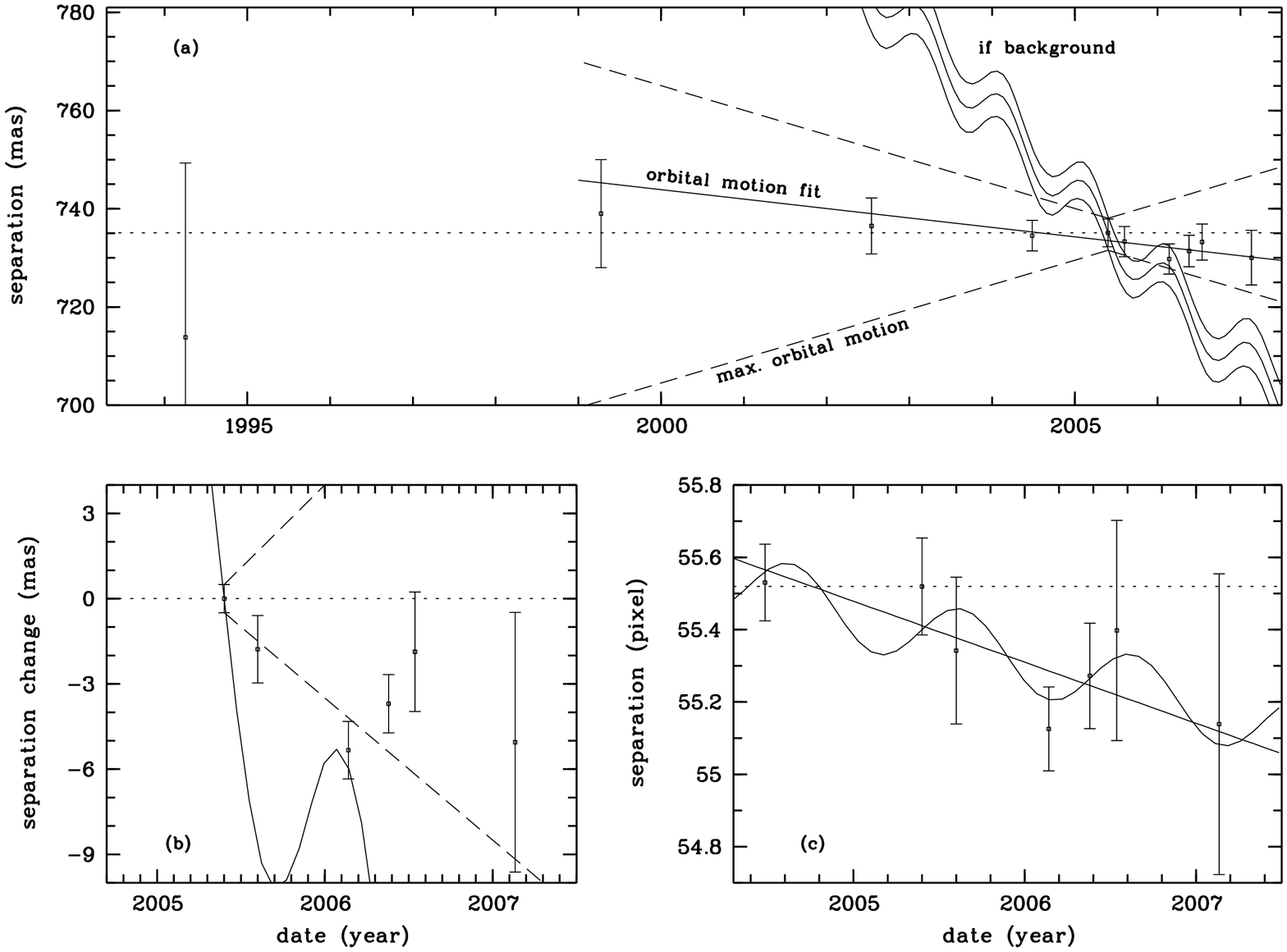}
\caption{Separation between GQ Lup A and its companion plotted versus time.
Data are from Janson et al. (2006) for 1994, N05 for 2002, and this paper for the rest.
Constant separation is indicated 
as a straight dotted line (through 2005.4 value), separation 
changes due to maximum 
expected orbital motion are indicated by dashed lines. 
All values are well within the expectations for orbital motion. 
(a) Absolute values from 1994 to 2007. 
Here, we also indicate as a wobbled line with
strongly decreasing values the expected separation change for the background
hypothesis, computed from the known GQ Lup A proper motion and parallax
and assuming that the faint object has negligible proper and parallactic motion
($\pm$ errors as expected from proper motion and parallax errors as 
additional wobbled lines);
the data are strongly deviant from the background hypothesis.
The full line from 1999 to 2007 with smaller negative slope is the best
fit for the orbital motion ($\sim 2$ to 3 mas/yr separation decrease from 
a linear fit to the last six data points), see text.
(b) Separation change since May 2005 with relative errors
(including the data point from the first new epoch 2005.4,
set to 0.0, with error bar just from Gaussian centering fits):
The data points from Feb and May 2006 are deviant from the dashed constant line 
(for constant separation) by 4.2 \& 2.6~$\sigma$, respectively, 
significance for orbital motion.
The wobbled line again is for the background hypothesis.
(c) Separation in pixel with absolute errors (since June 2004).
The full line with negative slope is our best fit for orbital motion as in (a);
this fit has a reduced $\chi ^{2}$ 
of 0.250 with 5 degrees of freedom and, hence, a probability of 0.94
(free fit parameters being starting separation and slope of decreasing separation).
The wobbled line here is for assuming that GQ Lup A and the faint object
would have different parallaxes (different by 1.5 mas as best fit);
this fit has a reduced $\chi ^{2}$ of 0.221 with 4 degrees of freedom and, hence,
a probability of only 0.92 
(free fit parameters being starting separation, slope of decreasing separation,
and difference in parallaxes between GQ Lup A and its co-moving companion).
Hence, the full line (for identical parallaxes for GQ Lup A and companion) is a 
better fit than the wobbled line. This gives evidence against the hypothesis
that the deviations in separations observed, in particular in 2005.6 and
2006.4 (compared to 2005.4), are due to different parallaxes for GQ Lup A and 
its companion.
}
\end{figure*}

Given the proper motion of GQ Lup A (see Mugrauer \& Neuh\"auser 2005),
we now have exceedingly high significance against the background
hypothesis (that the faint object is a non-moving background object
with negligible parallax; see Figs. 1 \& 2 and Table 2).
It is also very unlikely that both objects are independent members of 
the Lupus I cloud with similar proper motions (see Mugrauer \& Neuh\"auser 2005).

%

For final confirmation that they are bound,
one could, e.g., try to measure the parallaxes
of both objects with high precision (not yet feasible) or observe
curvature in the orbital motion (possible after a few more decades).

For $\sim 114$ AU projected separation (see Sect. 3.4) and a circular orbit, 
one would expect $\sim 1372$ years orbital period and, hence, up to a few
mas/yr orbital motion (e.g. up to 6.7 mas/yr for an eccentricity e=0.5
for edge-on inclination), and less than $0.3 ^{\circ}$ PA change for 
e=0 ($\le 0.82 ^{\circ}$ PA for e=0.5 for pole-on orbit).
The possible changes in separation and PA seen in Figs. 1 \& 2 are smaller
than these maxima, hence within the expectation.

Orbital motion is detectable as deviation among the separation and/or PA values.
The data points from Feb and May 2006 in the lower left panel (Fig. 2b)
are deviant from the May 2005 value (drawn as the full line for constant separation) 
by 4.2 \& 2.6~$\sigma$, respectively, so that we have a 
formal significance of 4.2 \& $2.6~\sigma$ for detection of orbital motion.
We observe a possible small (constant, linear) change in separation,
namely a decrease by $\sim 2$ to 3 mas/yr (best fit to the last six data points,
see Fig. 2), in particular a better fit compared to constant 
separation (732.1 mas being the mean) regarding rms scatter, which has a reduced
$\chi^{2} = 0.320$, 
and also a better fit compared to different parallaxes
for GQ Lup A and its companion.
We detect no significant change in PA.
This would be consistent with an orbital plane that is more edge-on than pole-on or
with a highly eccentric orbit, as already argued by
Janson et al. (2006) based on the data from 1994 until 2004 only.


The maximum motion of the GQ Lup companion of
$\sim 2$ to 3 mas/yr or $\sim 2$ km/s in one dimension is 
significantly smaller than the expected escape velocity from the GQ Lup A system, 
which is $5 \pm 2$ km/s, as already concluded with previous data 
in Mugrauer \& Neuh\"auser (2005).

From the measured projected equatorial rotational velocity $v \cdot \sin i = 6.8 \pm 0.4$ km/s,
the newly determined rotation period ($8.45 \pm 0.2$ days), and the luminosity 
and temperature of GQ Lup A, Broeg et al. (2007) estimated the inclination of the 
star (GQ Lup A) to be $27 \pm 5^{\circ}$ degrees, which is more pole-on than edge-on.
Unless the orbit of the companion is highly eccentric, this may indicate
that the stellar equator and the companion have different planes.
Normally, one would expect a planet to form in or nearly in the
equatorial plane of the star. If the GQ Lup companion is a planet,
however, it probably has not formed at its current wide separation,
but further inward, and was then moved to a larger separation by
an encounter with another, more massive inner protoplanet
(Debes \& Sigurdsson 2006; Boss 2006). 
An encounter with another star could have resulted in a dynamical perturbation
of the forming GQ Lup (system) resulting in a binary with large
mass ratio, i.e., a companion forming as a brown dwarf (embryo) 
by fragmentation, and/or would also have increased the
orbital inclination and/or eccentricity of the (forming)
companion significantly.

Deviations of the separation and PA values from being constant
could also be due to other reasons: \\
(i) If either GQ Lup A or b were an unresolved multiple,
one would expect a periodic variation in the separation.
If, e.g., GQ Lup A was an unresolved close binary made up
of two stars with different brightness, its photocenter would
show periodic variations according to the Keplerian motion of
the two stars around its common center-of-mass. Then, the
apparent separation between this photocenter and GQ Lup b would also vary
with the same period. Quantitatively, this effect depends on
the separation between the two stars (of this hypothetical binary)
and their brightness difference. If the observed change in separation
between GQ Lup A and b was due to this effect, the time-scale is
short, of the order of months to few years, so the separation between
the two stars would be limited to a few AU. Even for large mass and
brightness differences between the two stars, the effect would then
be limited to about 0.14 mas/yr. The observed effect is much larger 
($5.3 \pm 1.1$ mas/yr, comparing the separations in 2005.40 and 2006.14), 
hence cannot be explained by this effect. \\
(ii) GQ Lup A and its apparent companion b do not orbit each other,
but have slightly different proper motions. This possibility has
already been discussed in Mugrauer \& Neuh\"auser (2005).
Since signatures of accretion are detected in both objects
(A and companion; Seifahrt et al. 2007), both are young and, hence,
are most likely members of the Lupus I could. The velocity dispersion in star
forming regions like Lupus is only a few mas/yr, the same order of magnitude
as the difference in proper motion observed here.
The probability, however, to find two young objects within $\sim 0.7^{\prime \prime}$
with almost exactly the same proper motion (which is different from the
mean Lupus proper motion, Mugrauer \& Neuh\"auser 2005), 
and at the same time not orbiting each other, is exceedingly small. 
We could also show (in Fig. 2) that the assumption that the two objects
are unbound at different parallaxes gives a worse fit (to the observed
separation changes) than the assumption that they are at 
the same distance orbiting each other.
\\
(iii) The apparent effect could have been introduced by incorrect calibration.
We have corrected each original separation and PA measurement for GQ Lup
by the individual calibration results from HIP 73357. It could be that
pixel scale and/or detector orientation, however, do not change with time,
therefore the fact that we measure different values is not real, 
but just within the noise. However, when plotting 
separations in pixels (Fig. 2c) 
or the measured original un-corrected PA in degrees (not shown,
but see Table 2), i.e., both without astrometric calibration correction
versus time,
we still see the same effect, a small decrease in separation and 
no effect in PA (actually an even larger scatter in PA, 
so that the individual correction is most likely needed). \\
Hence, orbital motion is the most likely cause for the small
(linear) variation in the separation; quantitatively, the
variation is within expectation.

\subsection{The parallax of GQ Lup}

The NACO observations were also used to determine the parallax of GQ Lup
A and its companion.
The largest error source in the luminosity of the companion and, hence, 
its mass determination is the uncertain distance (90 to 190 pc for Lupus I, N05). 
In the north-west corner of the small (S13) NACO field-of-view (FoV), 
there is an additional star detected in all NACO images,
about 6.4$^{\prime \prime}$ NW of GQ Lup A. As any faint object near
a bright star, we have to see it as a companion candidate,
which we call GQ Lup/cc2 ({\em cc} for {\em c}ompanion {\em c}andidate,
GQ Lup b formerly was GQ Lup/cc1). 
See Table 3 for the separation of this object (relative to both GQ Lup A
and its confirmed companion b) corrected for pixel scale and detector orientation;
the largest contribution to the error budget is due to the large separation in pixels
(several hundred pixels) between the object cc2 and the two components
of the GQ Lup system, to be multiplied by the error in the pixel scale, 
and then taken into account in the proper error propagation.
The data from 2006.38 and 2007.13 have the largest errors, which are
due to the poor FWHM at these epochs (see Table 1), cc2 is only
marginally detected at those two epochs. We therefore exclude those two data 
points in the parallax determination (Table 3, Figs. 3 and 4), as they do 
not give any further constraints (they are consistent with the solution given).

\begin{table*}
\begin{tabular}{l|cc|cc}
\multicolumn{5}{c}{\bf Table 3. Astrometry on background star cc2 with VLT/NACO.} \\ \hline
Epoch   & \multicolumn{2}{c}{separation between GQ Lup A and cc2} & \multicolumn{2}{c}{separation between GQ Lup b and cc2} \\ 
year    & $\Delta RA$ [$^{\prime \prime}$] & $\Delta Dec$ [$^{\prime \prime}$] & $\Delta RA$ [$^{\prime \prime}$] & $\Delta Dec$ [$^{\prime \prime}$] \\ \hline
2004.48 & $4.3265 \pm 0.0164$ & $4.6473 \pm 0.0175$ & $3.5953 \pm 0.0137$ & $4.5775 \pm 0.0173$ \\
2005.40 & $4.3120 \pm 0.0164$ & $4.6739 \pm 0.0177$ & $3.5812 \pm 0.0136$ & $4.5969 \pm 0.0174$ \\
2005.60 & $4.2977 \pm 0.0163$ & $4.6715 \pm 0.0177$ & $3.5703 \pm 0.0135$ & $4.5956 \pm 0.0174$ \\
2006.14 & $4.3041 \pm 0.0163$ & $4.6909 \pm 0.0178$ & $3.5786 \pm 0.0136$ & $4.6134 \pm 0.0175$ \\
2006.54 & $4.2888 \pm 0.0163$ & $4.6965 \pm 0.0179$ & $3.5590 \pm 0.0136$ & $4.6272 \pm 0.0179$ \\ \hline
\end{tabular}
\end{table*}

We can first verify that GQ Lup/cc2 is a background star, Figs. 3 and 4,
because separation between this object and GQ Lup A 
(and also the separation between cc2 and the GQ Lup companion)
in both right ascension and declination change as expected from 
the known proper motion of GQ Lup A.
When assuming that cc2 does not move, then we obtain as proper motion
for GQ Lup A $\mu _{\alpha} = -16.8 \pm 4.2$ mas/yr 
and $\mu _{\delta} = -24.3 \pm 2.2$ mas/yr,
and for the known co-moving GQ Lup companion 
we get $\mu _{\alpha} = -15.2 \pm 4.6$ mas/yr and 
$\mu _{\delta} = -23.7 \pm 2.7$ mar/yr.\footnote{These two proper motions are 
not only consistent with each other,
but also consistent with the published proper motion of GQ Lup A being
$\mu _{\alpha} = -19.15 \pm 1.67$ mas/yr and $\mu _{\delta} = -21.06 \pm 1.69$ mas/yr 
(Mugrauer \& Neuh\"auser 2005).}
Hence, the (proper and parallactic) motion of cc2 is negligible, 
therefore we can use it
for measuring the parallax of GQ Lup A and its companion from the scatter in 
Figs. 3 and 4, namely deviations from a linear motion. 
This background object (GQ Lup/cc2) is listed in USNO-B1 as object 00543-0373323
with B$_{2}$ = 19.28 mag, R$_{2}$ = 17.77 mag, and I = 15.94 mag;
it is also listed in NOMAD as object 0543-0380757 with B = 19.53 mag.
We have determined K$_{\rm s}$ = $15.01 \pm 0.33$ mag (see Table 4).
Hence, it is probably a mid- to late-K dwarf (or giant) star with weak extinction.
If it is on the main sequence, it has a distance of 800 to 1700 pc,
i.e., a parallax of $1.0 \pm 0.3$ mas (even smaller if a giant). 
Its own proper motion is negligible, as seen above.

The timing of our NACO observations in Feb, May, and Aug 2005 to
2007 was chosen, because the offset in the separation due to 
parallactic motion is maximal at these times.

The wobble remaining in the motion of GQ Lup A (and its companion)
after subtracting the proper motion is then the parallactic motion.
In Figs. 3 and 4, we show the data points (difference in position between
GQ Lup A and companion, respectively, and the third star cc2), and our best fit 
giving the parallax of GQ Lup A to be $7.4 \pm 1.9$ mas
and the parallax of the GQ Lup companion to be $8.2 \pm 2.1$ mas
(full error propagation).
For the fit to the GQ Lup A data, the wobbled curve with a non-zero
parallax has a reduced $\chi^{2} = 0.036$ for RA and
$\chi^{2} = 0.011$ for Dec, while a linear fit (decreasing separation, 
no parallax) has $\chi^{2} = 0.167$ for RA and 0.037 for Dec.
Hence, our fit with $\pi = 7.4 \pm 1.9$ mas is better.
For the fit to the data on the GQ Lup companion, the wobbled curve with a non-zero
parallax has a reduced $\chi^{2} = 0.022$ (for RA) and 0.039 (for Dec), 
while a linear fit (decreasing separation, no parallax) has $\chi^{2} = 0.288$
(for RA) and 0.058 (for Dec).
Hence, again, our fit with $\pi = 8.2 \pm 2.1$ mas is better.

The two values for GQ Lup A and its companion are consistent with each other.
For GQ Lup A, the parallax corresponds to $\sim 135$ pc (110 to 180 pc) 
or roughly $135 \pm 30$ pc, i.e., more precise
than the previous distance estimate for the Lupus I cloud
including GQ Lup being $140 \pm 50$ pc (N05).
For the GQ Lup companion, the parallax corresponds to $\sim 122$ pc (100 to 160 pc).

Our parallax measurements are based on only one comparison star (cc2), 
which could move itself (proper and/or parallactic motion).
The above values are correct when assuming that this background 
object has negligible proper and parallactic motion.
The fact that we obtain the correct GQ Lup proper motion 
when comparing to this comparison star shows that
its proper motion is very small.

The error in the parallax determination is quite large,
which is due to the following reasons. \\
(i) We use only one comparison star. \\
(ii) The GQ Lup companion and the comparison star are quite faint. \\
(iii) There are statical abberrations and field distortions
of the CONICA S13 camera amounting to up to 3 mas at its edges.\footnote{We
are monitoring these field distortions in a different 
(astrometric) observing program, where we observe the 47 Tuc
cluster center several times per year with the NACO S13 FoV
and always the same set-up (also 2005 to 2007 as GQ Lup). 
We achieved sub-mas relative precision; first results are 
published in Neuh\"auser et al. (2007) and Seifahrt et al. (2007),
more details will be given in Roell et al. (in preparation). 
From these observations, we already know that the field 
distortions are very small except at the field edges.}
Since we measure relative changes between different epochs,
not absolute positions, the important question for a parallax
measurement is the repeatability of the configuration between the
epochs. Since the jitter/dither pattern at each epoch moves the
reference line (GQ Lup to field star) over the FoV, the variations
in this separation from the field distortion are dominating the
uncertainties in the angle and length of the reference line in
each epoch and are contained in the error budget. 
Since we have always chosen a small jitter/dither box/offset,
GQ Lup and its very nearby companion are always located near
the center of the FoV and the comparison star is also always 
located nearly at the same spot on the detector (within few arc sec),
so that the distortions seen by the camera are always the
same between the epochs. Since this is a relative measurement, not
an absolute one, the remaining distortion term affecting the true
length and angle of the reference line is not important here. \\
(iv) Differential chromatic refraction has a negligible effect in the K band.
The wavelength dependence of the refractive index of air is very low in K,
and the differential terms from the color difference of GQ Lup and 
the background star are much smaller (both are late K-type dwarfs),
because we operate in the Rayleigh-Jeans term of the spectral
energy distribution of both targets. 
The remaining effect is much smaller than 0.5 mas. \\
(v) Differential AO correction is certainly an issue when
comparing GQ Lup or its companion to the comparison star,
because it is located several arc sec away, while GQ Lup and
its companion, within 1 arc sec, are inside the isoplanatic angle. \\
Effects (i), (ii), and (v) are largest and lead to the
relatively large error bar. However, the parallactic wobble
effect is still clearly seen in Figs. 3 \& 4.

What we measure is the difference between the GQ Lup parallax 
and the parallax of this comparison star. If we further assume that 
it is a background star at large distance with negligible parallax,
then we can interpret the measurement as the parallax of GQ Lup.
We can also subtract the probable parallax of the background star
($1.0 \pm 0.3$ mas, see above) from the total difference in 
parallax measured here, thus obtaining $6.4 \pm 1.9$ mas as parallax 
of GQ Lup A (i.e. $156 \pm 50$ pc) and also
$7.2 \pm 2.1$ mas as parallax of GQ Lup b (i.e. $139 \pm 45$ pc).
These values are not more precise than the original assumption
($140 \pm 50$ pc, N05 for the Lupus I cloud), but our
measurements show that GQ Lup A and its companion are 
located at the same distance and at the same distance
as the Lupus I cloud and its members.

Given these uncertainties and limitations, our measurement needs 
to be confirmed. If correct, it may be the smallest parallax determined
from the ground
(but with large error).
This method can also be applied to other suitable targets, 
which have a background object within the small AO FoV.

\begin{figure}
\centering
\includegraphics[width=70mm]{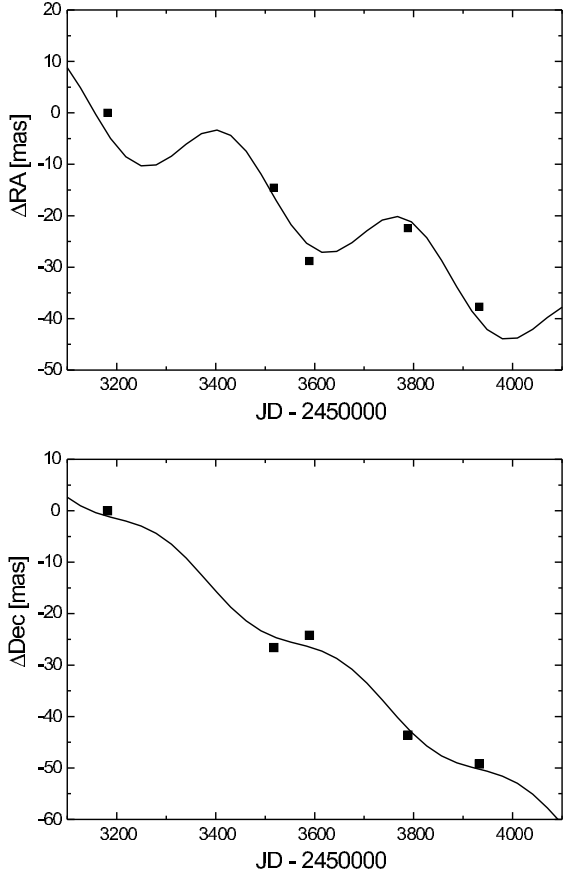}
\caption{Proper and parallactic motion of GQ Lup A 
in right ascension (top) and declination (bottom) relative to the
background object GQ Lup/cc2.
Changes in separations in mas (since 2004.5) are plotted versus observing epoch as Julian date JD.
We also show our best fit yielding 
a parallactic wobble of $7.4 \pm 1.9$ mas, from which we have to subtract the
probable parallax of the background object cc 2 ($1.0 \pm 0.3$ mas, see text),
thus obtaining $6.4 \pm 1.9$ mas or $156 \pm 50$ pc for GQ Lup A.}
\end{figure}

\begin{figure}
\centering
\includegraphics[width=70mm]{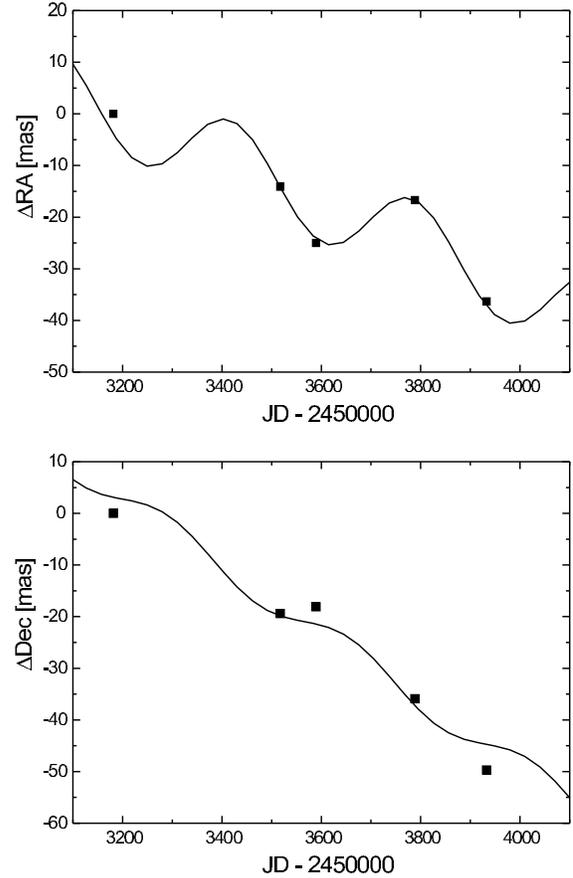}
\caption{Proper and parallactic motion of GQ Lup b
in right ascension (top) and declination (bottom) relative to 
the background object GQ Lup/cc2.
Changes in separations in mas (since 2004.5) are plotted versus observing epoch as Julian date JD.
We also show our best fit yielding a parallactic wobble of
$8.2 \pm 2.1$ mas, from which we have to subtract the
probable parallax of the background object cc2 ($1.0 \pm 0.3$ mas, see text),
thus obtaining $7.2 \pm 2.1$ mas or $139 \pm 45$ pc for 
the GQ Lup companion.}
\end{figure}

\section{Photometry}

During the service mode observations of GQ Lup, it was either
clear at least in and around Lupus and around HIP 73357 or photometric;
HIP 73357 was always observed either immediately before or after GQ Lup.
After aperture photometry on HIP 73357 A and B, we notice
that the difference between the K$_{\rm s}$-band magnitudes of A and B does
not vary with time (see Fig. 5 bottom panel). Hence, they can both
be assumed to be constant (they are neither known 
nor expected to be variable). We can then use their 2MASS magnitudes
(K$_{\rm s} = 6.467 \pm 0.016$ mag for HIP 73357 A
and K$_{\rm s} = 8.027 \pm 0.021$ mag for HIP 73357 B)
and the respective airmasses to correct the instrumental magnitudes 
obtained for the objects in the GQ Lup field.
Aperture photometry was performed in the same way on GQ Lup A, 
its companion (after subtraction of PSF of GQ Lup A), 
and the background object cc2. See Table 4 for the K$_{\rm s}$-band magnitudes.
The error budget includes the 2MASS magnitude errors of HIP 73357 A and B,
the measurement errors in aperture photometry, airmass, and extinction coefficient.
We display the variation of K$_{\rm s}$ band magnitudes with time in Fig. 5
for the difference between HIP 73357 A and B, GQ Lup A, its companion b, 
and the background object cc2.
 
\begin{table}
\begin{tabular}{lcccc}
\multicolumn{5}{c}{\bf Table 4. Photometry on GQ Lup with VLT/NACO.} \\ \hline
Epoch   & diff. (a) & \multicolumn{3}{c}{K$_{\rm s}$ band magnitude [mag]} \\ 
year    & HIP A - B & GQ Lup A & GQ Lup b & cc2 (b) \\ \hline
2005.40 & $1.56 \pm 0.07$ & $7.07 \pm 0.07$ & $13.46 \pm 0.12$ & $14.80 \pm 0.24$ \\
2005.60 & $1.46 \pm 0.13$ & $7.04 \pm 0.14$ & $13.34 \pm 0.42$ & $14.78 \pm 0.18$ \\
2006.14 & $1.57 \pm 0.08$ & $7.09 \pm 0.08$ & $13.38 \pm 0.10$ & $14.88 \pm 0.16$ \\
2006.38 & $1.59 \pm 0.07$ & $7.40 \pm 0.08$ & $13.47 \pm 0.14$ & $14.87 \pm 0.26$ \\
2006.54 & $1.66 \pm 0.12$ & $7.02 \pm 0.13$ & $13.45 \pm 0.38$ & $15.07 \pm 0.34$ \\
2007.13 & $1.54 \pm 0.16$ & $7.48 \pm 0.16$ & $13.27 \pm 0.12$ & $15.67 \pm 0.92$ \\ \hline
mean    & $1.56 \pm 0.07$ & $7.18 \pm 0.20$ & $13.39 \pm 0.08$ & $15.01 \pm 0.33$ \\ \hline
\end{tabular}

Remarks: (a) difference in K$_{\rm s}$ between HIP 73357 A and B.
(b) background object GQ Lup/cc2.
\end{table}

The mean magnitude of GQ Lup A (Table 4) is consistent with its
2MASS value (K$_{\rm s} = 7.096 \pm 0.020$ mag), obtained at an unknown rotation phase.
We confirm the variability of GQ Lup A in the K$_{\rm s}$-band ($4~\sigma$ significance)
already found by Broeg et al. (2007), interpreted as being due to surface
spots and a $\sim 8.4$ day rotation period. Broeg et al. (2007)
found an amplitude in K$_{\rm s}$ in April/May 2005 of $\pm 0.22$ mag; 
we obtain $\pm 0.20$ for May 2005 to Feb 2007.
In our data from May 2005 to Feb 2007, GQ Lup A varies between 7.02
and 7.48 mag (Feb 2007), while it varies between 6.85 and 7.28 mag in April 2005 
(Broeg et al. 2007 and C. Broeg, private communication).
There may be a slight dimming of GQ Lup A with time,
possibly due to temporal changes in the spottedness.

The mean magnitude of the GQ Lup companion given here (Table 4) 
is slightly fainter than given in N05 ($13.10 \pm 0.15$ mag for June 2004), 
but within $2~\sigma$ errors, due to the fact that we have now 
improved on the subtraction of the GQ Lup A PSF (see above).
The new mean value is consistent with the value given by Marois
et al. (2007) for the 2002 Subaru exposure (K$_{\rm s} = 13.37 \pm 0.12$ mag).
Given its faintness and the large error bars, we cannot find evidence for 
variability of the GQ Lup companion in the K$_{\rm s}$-band. The K$_{\rm s}$-band values show a small
standard deviation of $\pm 0.08$ mag, which could be due to variability
in the GQ Lup companion and/or HIP 73357;
possible variations would be expected to have a smaller amplitude than 
in GQ Lup A, just because the companion is much fainter. Variability would be expected, 
given that it is a young object ($\le 2$ Myr as GQ Lup A) and that 
Paschen $\beta$ was found to be in emission by Seifahrt et al. (2007), 
a sign of ongoing accretion.

\begin{figure*}
\centering
\includegraphics[width=110mm,angle=270]{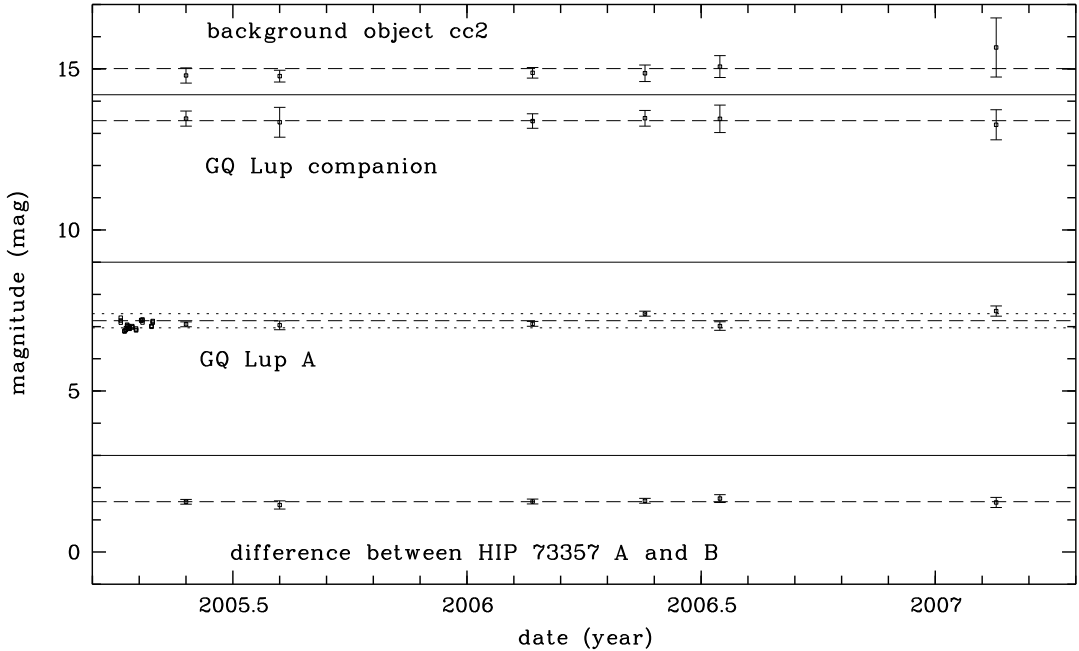}
\caption{K$_{\rm s}$ band magnitude versus time
for difference between HIP 73357 A and B (bottom panel),
GQ Lup A (2nd from bottom), GQ Lup companion (2nd from top),
and the background object GQ Lup/cc2 (top panel).
We always show as a long-dashed line the mean value.
In the case of GQ Lup A, we also show the data points from Broeg et al. (2007),
for which absolute photometry is available (C. Broeg, priv. comm.)
as small dots in April 2005; we also show as dotted lines the
range in variability ($\pm 0.22$ mag) found by Broeg et al. (2007)
during their monitoring of GQ Lup A (yielding the rotation period
to be $\sim 8.4$ days). 
We find variability in GQ Lup A (amplitude $\pm 0.20$ mag, 4~$\sigma$).}
\end{figure*}

\section{Deep imaging}

We use all imaging data obtained from June 2004 to July 2006
(see Table 1, i.e., omitting the data from February 2007 because of low quality,
e.g., strong reflections and waffle structure) to combine all of
them together by shift+add (as usual with ESO eclipse) to get a 
very deep image with very high dynamic range. 
We use the IDL routine for PSF subtraction as explained above
to subtract the PSF of GQ Lup A. The resulting image after
PSF subtraction is shown in Fig. 6. We can then measure the
background level in all pixels and at all separations from the
(former) photocenter of GQ Lup A. 

\begin{figure}
\centering
\includegraphics[width=65mm,angle=0]{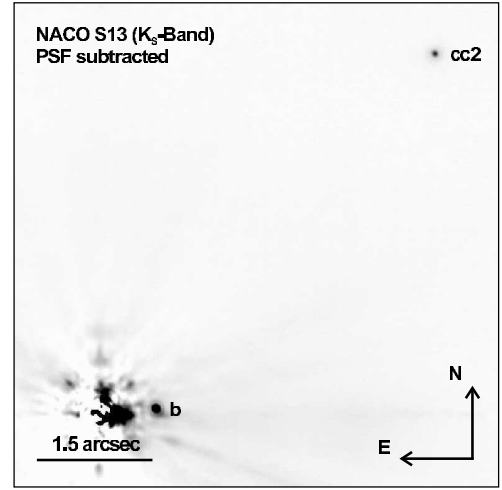}
\caption{Deepest image of GQ Lup (102 min total exposure)
after subtraction of PSF of GQ Lup A. North is up, east to the left.
The companion GQ Lup b is seen well 0.7$^{\prime \prime}$ west of GQ Lup A,
whose PSF was subtracted. The background object GQ Lup/cc2 is seen
in the upper right (NW) corner. }
\end{figure}

\begin{figure}
\centering
\includegraphics[width=65mm,angle=270]{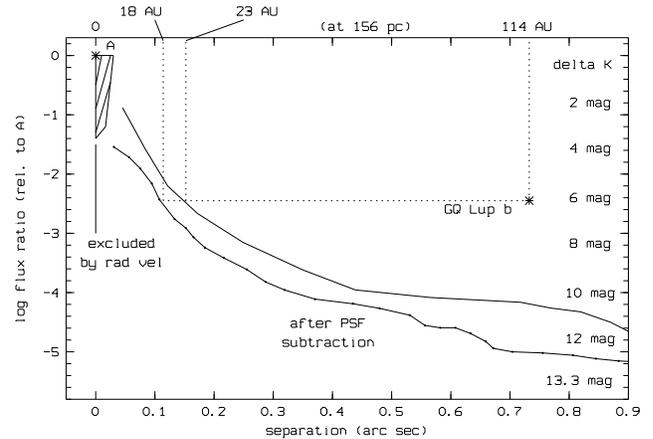}
\caption{Dynamic range from deepest image shown in Fig. 6.
We plot (log of) noise flux (S/N=3) devided by the peak flux
of GQ Lup A (before PSF subtraction) on the left axis (and the magnitude difference
at the right axis) versus separation in arc sec on the bottom axis
(versus separation in AU at 156 pc on the top axis).
The lower full line with points is obtained after PSF subtraction,
the upper line is without PSF subtraction. 
GQ Lup and its companion are indicated 
as star symbols.
The hatched area in the upper left is the region in this parameter space,
where companions can be excluded by high-resolution spectroscopic 
monitoring of the radial velocity (rad vel) by Broeg et al. (2007), see text. 
}
\end{figure}

The flux ratio between the
peak of GQ Lup A (before PSF subtraction) and the background
(in the PSF-subtracted image) versus the separation is then
plotted in Fig. 7 and compared to the dynamic range before
PSF subtraction. No further companion candidates are detected.
We can exclude companion candidates with at least the flux
of GQ Lup b outside of $\sim 0.115^{\prime \prime}$ or $\sim 18$ AU (at 156 pc),
or outside of 23 AU without PSF subtraction.
Closer companions (or those located before or behind GQ Lup A)
cannot be excluded from the imaging alone.
As displayed in Fig. 7, we would be able to detect companions, 
e.g., $\le 10$ mag fainter (in K$_{\rm s}$) than GQ Lup A at $\ge 0.3^{\prime \prime}$ 
separation or $\le 12$ mag fainter at $\ge 0.7^{\prime \prime}$ (after PSF subtraction).
Between about 0.6$^{\prime \prime}$ and 0.9$^{\prime \prime}$, we could 
gain $\sim 1$ mag dynamic range by PSF subtraction.

Broeg et al. (2007) have monitored GQ Lup for several years with the ESO high-resolution
spectrographs FEROS and HARPS and searched for companions with the
radial velocity method; they could not detect any spectroscopic companions
with mass above $\sim 0.1$~M$_{\odot}$ inside $\sim 2.6$ AU.
With L$_{\rm bol} \simeq 0.1$~L$_{\odot}$ for a few Myr young $\sim 0.1$~M$_{\odot}$
star (Burrows et al. 1997) and using L$_{\rm bol} = 1.58$~L$_{\odot}$ for GQ Lup A (N05), 
we can also display the area in Fig. 7,
where companions can be excluded by the spectroscopic monitoring.
This area is complementary to the area investigated by the imaging.

Debes \& Sigurdsson (2006) and Boss (2006) suggested that GQ Lup b,
if a planet, would have been ejected by a near encounter with another
protoplanet, which would most likely be located within $\sim 10$ AU
of GQ Lup A and would need to be more massive than GQ Lup b.
We could not detect such a close-in massive planet, yet.
The radial velocity monitoring of GQ Lup A (Broeg et al. 2007) also could not 
yet detect a close-in massive planet.
However, there is still a large separation range not yet probed
with either imaging or spectroscopy (between few and 18 AU), 
where further companions can hide.

\section{Summary of results and conclusions}

Given the newly constrained mean K$_{\rm s}$-band magnitude of $13.39 \pm 0.08$ mag 
and distance measured for the companion of $139 \pm 45$ pc (this paper)
and with B.C.$_{\rm K}= 3.0 \pm 0.1$ mag (following Golimowski et al. 2004)
and the temperature of the companion newly constrained in Seifahrt et al. (2007), 
we can re-estimate the luminosity of the companion to be 
$\log (L_{\rm bol}/L_{\odot}) = -2.375 \pm 0.245$, 
similar to the value in N05, but with a smaller error bar.

With the temperature $2650 \pm 100$~K and gravity
$\log g = 3.7 \pm 0.5$ dex (g in $g/cm^{2}$) for the GQ Lup companion (Seifahrt et al. 2007), 
we can use luminosity and temperature to re-calculate its radius 
to be $3.0 \pm 0.5$ R$_{\rm Jup}$.
With radius and gravity, we obtain a mass of $\sim 20$ M$_{\rm Jup}$
with a possible minimum (value $- 1 \sigma$) being only few M$_{\rm Jup}$,
and the maximum being around the sub-stellar limit.
However, the upper mass limit for the GQ Lup companion is still $36 \pm 3$~M$_{\rm Jup}$, 
because the GQ Lup companion is smaller, cooler, and fainter than both components 
in the eclipsing double-lined spectroscopic binary brown dwarf 2M0535 
(Stassun et al. 2006), which has a similar age as GQ Lup,
as already noticed by Seifahrt et al. (2007).

We can also use luminosity L, temperature T, gravity g, radius R, and the age of the 
young T Tauri star GQ Lup ($\le 2$ Myr, N05, having strong IR excess) to estimate 
the mass of the companion from theoretical evolutionary models
(as done in N05 with the original, less constrained parameters):
from Burrows et al. (1997), we consistently obtain for all combinations of L, T, g, R, and age
a mass of $\sim 20$ M$_{\rm Jup}$, and from Baraffe et al. (2002), we consistently 
get for all combinations L, T, and age a mass of $\sim 20$ M$_{\rm Jup}$. 
According to the calculations following Wuchterl \& Tscharnuter (2003), as plotted in 
Fig. 4 in N05, the GQ Lup companion would have $\sim 5$ M$_{\rm Jup}$.
It may be seen as intriguing that both the atmospheric and the conventional evolutionary 
models consistently give $\sim 20$ M$_{\rm Jup}$ as the best value.
However, we note that the models by Burrows et al. and Baraffe et al. may not be 
valid for very young objects ($\le 10$ Myr), as initial conditions are not taken into account,
and none of the models used are tested positively for very low-mass objects or calibrated.

A better mass estimate can be obtained in the future by comparison with more very young objects
with dynamically determined masses and/or atmospheric or evolutionary models that are calibrated.

We can summarize our results as follows.
\begin{enumerate}
\item With precise relative astrometry at six new epochs, we have detected small 
deviations in the separation between GQ Lup A and its companion, consistent with
orbital motion of $\sim 2$ to 3 mas/yr.
\item It remains unclear whether the slight decrease in separation
observed is due to orbital motion around each other (bound) or due to different 
parallaxes (unbound). The latter case is, however, less likely.
\item By comparing the position of GQ Lup A and its companion
to a third object in the small NACO FoV, a background star at 800 to 1700 pc,
we could determine the parallaxes of GQ Lup A and its companion
corresponding to a distance of $156 \pm 50$ pc for A and $139 \pm 45$ pc for
the companion.
\item Apart from that background star, no additional companion candidates are 
detected. Outside of 115 mas (18 AU), we can exclude further companion candidates 
as bright or brighter than the known co-moving companion.
\item We could confirm photometric variability of GQ Lup A in the K$_{\rm s}$-band (7 exposures
over 3 years); variability of the companion is smaller than $\pm 0.08$ mag.
\item From the newly constrained distance and the mean K$_{\rm s}$-band magnitude, we re-estimate 
luminosity, radius, and mass of the companion, 
but we can still not decide whether the companion is below the deuterium burning mass limit 
(or below the radial velocity brown dwarf desert, which is 
at $\sim 30$ M$_{\rm Jup}$ according to Grether \& Lineweaver 2006), 
i.e., whether the companion is of planetary mass or a brown dwarf.
The large uncertainty in mass is due to uncertainties in the theoretical
models and the gravity and distance measurements.
\end{enumerate}

\begin{acknowledgements}
We are gratefull to ESO's User Support Department for help in preparation
of the service mode observations and carrying out our observations.
TOBS would like to thank Evangelisches Studienwerk e.V. Villigst for financial support.
NV acknowledges support by FONDECYT grant 1061199.
We would like to thank Christopher Broeg for providing his photometric 
data on GQ Lup A in electronic form.
\end{acknowledgements}

\begin{appendix} 

\section{Astrometric calibration}

For the astrometric calibration, we observed from 2005 to 2007 always the
same binary star HIP 73357, where the Hipparcos satellite measured the
separation to be
$8.43 \pm 0.03^{\prime \prime}$ and
the position angle (PA, measured from north over east to south)
to be $337.30 \pm 0.06^{\circ}$ (at epoch 1991.25).
HIP 73357 A and B have spectral types mid-A and mid-F,
hence masses of $\sim 2.5$ and $\sim 1.5$~M$_{\odot}$, and a distance of $104 \pm 22$ pc.
This results in an orbital period of $\sim 13000$ years (for circular orbit,
shorter if eccentric) and, hence, a maximum orbital motion 
of $\sim 2.5$ mas/yr for edge-on inclination
and $\sim 0.025^{\circ}$/yr for face-on inclination 
(both for an eccentricity of $e \simeq 0.2$).\footnote{We confirm a-posteriori, 
see below, that these assumptions are probably correct. In reality, we 
have started with slightly other assumptions and have then iterated;
in the iterations, the values for pixel scale and detector orientations stayed
constant, just the (maximal) errors due to (maximal) orbital motion changed.}
It is not confirmed that HIP 73357 A and B form a bound pair
orbiting each other (curvature in orbital motion is not yet detected). 
Given their known (and very similar) proper motions (they form
a common-proper-motion pair), the motion of A relative to B 
results in slightly less changes in separation and PA than the 
maximum orbital motion given above.

Absolute values and their errors for
separation and PA measured for HIP 73357 and GQ Lup images include 
the errors in separation and PA for HIP 73357 AB from Hipparcos,
errors from Gaussian centering fits
on HIP 73357 A and B, and GQ Lup A and companion,
as well as the
maximum possible (orbital) motion (of HIP 73357 A relative to B)
between 1991.25 and the new observation at each epoch.
For absolute errors (given in Table 1) we include the maximum
(orbital) motion from 1991.25 to the new epoch. 
Values for changes in separation and PA with their (relative) errors 
(as given in Table 1) include only the maximum relative (orbital) 
motion since the first new epoch (2005.4, we observed a different 
astrometric calibration binary in 2004) and, of course, also the
errors from Gaussian centering fits.
The relative errors are smaller than the absolute errors.
This way, we can obtain better precision to detect orbital motion
in GQ Lup as changes in separation and/or PA, but we cannot give 
anymore the absolute values for separation (in arc sec) or PA on sky 
(in degrees), but can measure their {\em changes}.

The detector orientations given in Table 1 are the difference between 
the observed PA and 337.3$^{\circ}$ (from Hipparcos at 1991.25).
To obtain the true PA of, e.g., the GQ Lup companion,
we have to add the PA value measured on GQ Lup images
with the value given for the detector orientation.
Such an absolute value for the PA 
(similar for pixel scale and separation)
can also be given with (absolute) errors (accuracy). 
The relative errors are not applicable to the absolute
PA values, but only to PA changes since 2005.4 (relative errors for precision).

The values for separation and PA at 2005.4 have also their own
measurement errors (from Gaussian centering only, without uncertainty in 
(orbital) motion of HIP 73357), also listed in Table 2 under relative error.

\begin{figure}
\centering
\includegraphics[width=65mm,angle=270]{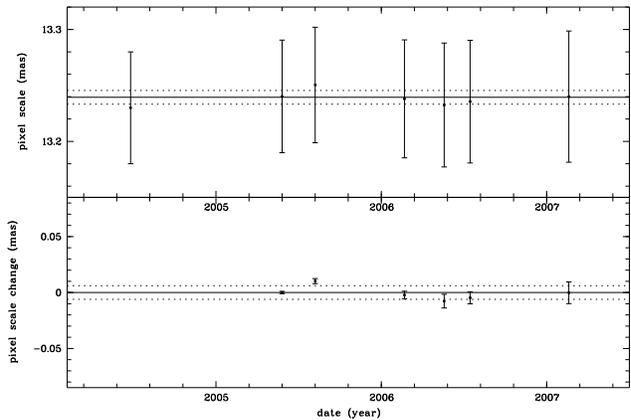}
\caption{NACO pixel scales as determined from binary HIP 73357 
observed within one hour from GQ Lup (but see footnote 1 for Feb 2007).
We show absolute values and errors
in the upper panel (including errors due to (orbital) motion in 
HIP 73357 since Hipparcos at epoch 1991.25) and relative values
for pixel scale changes with relative errors in the bottom panel
(including errors due to (orbital) motion in HIP 73357 only
since our first new epoch 2005.4),
and also showing the data point at the first new epoch 2005.4,
set to 0.0, with its error bar just from the Gaussian centering fit. 
Both panels show the same
range in pixel scale on the y-axis, 0.17 mas in both panels.}
\end{figure}

\begin{figure}
\centering
\includegraphics[width=65mm,angle=270]{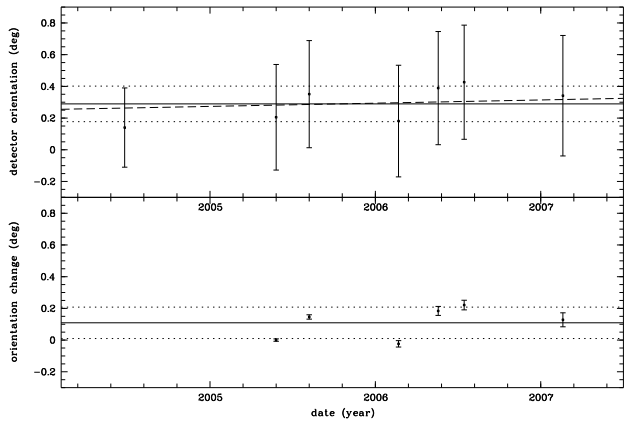}
\caption{NACO detector orientations as determined from binary HIP 73357
observed within one hour from GQ Lup (but see footnote 1 for Feb 2007).
Upper and lower panels are as in fig. A.1. The dashed line in the upper panel shows
the best fit to the data consistent with a detector orientation of $0^{\circ}$
at 1991.25, the Hipparcos epoch, indicating that the small increase in detector 
orientation with time can be interpreted as orbital motion in HIP 73357 A and B. 
Both panels show the same range in PA on the y-axis, $1.2^{\circ}$ in both panels.}
\end{figure}

We display the pixel scale and detector orientation values for all epochs
in Figs. A.1 and A.2. The mean pixel scale is $13.2394 \pm 0.0061$ mas/pixel
with 0.054 mas as the mean absolute error bar and 0.0031 mas as the mean relative error.
The detector orientation has a mean of $0.29 \pm 0.11^{\circ}$;
for orientation {\em changes} since May 2005, the mean is $0.09 \pm 0.11^{\circ}$
(and $0.024^{\circ}$ as mean relative error).
These values may indicate that HIP 73357 A and B show
an orbital motion of $\sim 0.3^{\circ}$ since 1991.25
(similar as assumed, see footnote 4), but we would need absolute
astrometric reference objects as a confirmation.
If true, the precision of the NACO detector alignment during 
target acquisition is around $\pm 0.1$ degree. 
Fitting the seven data points shown in Fig. A.2 (upper panel) with a constant 
line without slope (mean detector orientation) gives a similarily good fit 
(i.e. larger rms, reduced $\chi ^{2} = 0.85$) 
as fitting it to a detector orientation increasing constantly 
from $0^{\circ}$ at 1991.25 to the mean value 0.29 observed at the mean
of the new observing epochs (2005.9536), as drawn in Fig. A.2 (reduced $\chi^{2} = 0.97$).
The pixel scales determined do not show any slope (Fig. A.1).

The observed orbital motion in HIP 73357 is near the maximum orbital motion  
estimated above ($\sim 0.39^{\circ}$ from 1991.25 to 2005.4), 
therefore (i) the orbital inclination may well be near pole-on,
(ii) the eccentricity near $\sim 0.2$, and (iii) the change in separation should
be negligible, which we see in the roughly constant pixel scale obtained by
assuming that the separation between HIP 73357 A and B is constant.

The large change in PA in HIP 73357 results in larger errors in 
the PA calibration,
while the errors in pixel scale calibration remain very small, 
because the separation remains (nearly) constant;
absolute PA errors are always larger than $0.3^{\circ}$, the offset between 
the Hipparcos value and as observed with NACO, the difference
between $0.3^{\circ}$ and the PA errors given in Table 2
(even at epoch 2005.4) are due to Gaussian centering errors; 
errors in both PA and pixel scale - with few exceptions - increase with 
time due to additional possible orbital motion.

The stars HIP 73357 A and B are located near the edges
of the small S13 NACO FoV, and the jitter offsets are relatively
large in right ascension, so that field distortions may play an important
role. In an attempt to check the errors of our astrometry,
we not only measured separations and orientations in one full final high-S/N image
after combining all images of one epoch (one night), but we also 
measured separation and PA between HIP 73357 A and B (and similar for GQ Lup),
in any of the five HIP 73357 images (43 good S/N images of
HIP 73357 for Feb 2007, see footnote 1).
We can then take the mean of the separation and PA values and their
standard deviation as alternative measurement.
They are in agreement with the standard measurement (obtained after
combining all five images of HIP 73357) within small errors.
In the case of HIP 73357, the errors of these two measurement methods
are comparable. To be conservative, we always use the larger error 
value for absolute errors (as listed in Tables 1 \& 2).

\end{appendix}


\begin{thebibliography}{}

\bibitem[bar]{bar02} Baraffe I., Chabrier G., Allard F., Hauschildt P.H. 2002, A\&A 382, 563

\bibitem[boss]{boss06} Boss A.P., 2006, ApJ 637, L137

\bibitem[broeg]{broeg07} Broeg C., Schmidt T.O.B., Guenther E.W., Gaedke A., 
Bedalov A., Neuh\"auser R., Walter F.M., 2007, A\&A 468, 1039

\bibitem[bur97]{burr97} Burrows A., Marley M., Hubbard W. et al. 1997, ApJ 491, 856

\bibitem[chabr]{ch05} Chabrier G., Baraffe I., Allard F., Hauschildt P.H., 2005,
{\it Review on low-mass stars and brown dwarfs}. In: Resolved Stellar Populations
Conf. Proc. (arXiv:astro-ph/0509798)

\bibitem[debes]{debsig06} Debes J.H., Sigurdsson S., 2006, A\&A 451, 351

\bibitem[goli]{goli04} Golimowski D.A., Leggett S., Marley M. et al. 2004, AJ 127, 3516

\bibitem[janson]{jans06} Janson M., Brandner W., Henning T., Zinnecker H., 2006, A\&A 453, 609

\bibitem[grelin]{gre06} Grether D., Lineweaver C.H., 2006, ApJ 640, 1051

\bibitem[krikp]{kir06} Kirkpatrick J.D., Barman T.S., Burgasser A.J., et al., 2006, ApJ 639, 1120

\bibitem[marois]{mar07} Marois C., Macintosh B., Barman T., 2007, ApJ 654, L151	

\bibitem[mcel]{mcelw07} McElwain M.W., Metchev S.A., Larkin J.E., et al., 2007, ApJ 656, 505	

\bibitem[mugrauer05]{mugi05} Mugrauer M., Neuh\"auser R., 2005, AN 361, L15

\bibitem[neuh05]{neuh05} Neuh\"auser R., Guenther E.W., Wuchterl G., Mugrauer M.,
Bedalov A., Hauschildt P., 2005, A\&A 435, L13 (N05)

\bibitem[neuh07]{neu07} Neuh\"auser R., Seifahrt A., Roell T., Bedalov A., Mugrauer M., 2007,
`Detectability of Planets in Wide Binaries by Ground-Based Relative Astrometry with AO'.
In: Hartkopf W.I., Guinan E.F., Harmanec (Eds.) IAU Symp. 240 on
Binary Stars as Critical Tools and Tests in Contemporary Astrophysics,
pp. 261-263, arXiv:astro-ph/0610547

\bibitem[rous]{rous03} Rousset G., Lacombe F., Puget P., et al., 2003, SPIE 4839, 140

\bibitem[seifa]{seif05} Seifahrt A., Neuh\"auser R., Hauschildt P.H., 2007, A\&A 463, 309

\bibitem[seif]{seif07} Seifahrt A., Roell T., Neuh\"auser R., 2007, 
`Probing micro-arcsec astrometry with NACO',
In: Proceedings of the 2007 ESO Instrument Calibration Workshop, 
in press, arXiv:astro-ph/0706.2613

\bibitem[stass]{stass06} Stassun K.G., Mathieu R.D., Valenti J.A., 2006, Nature 440, 311	

\bibitem[wu03]{wuchterl03} Wuchterl G., Tscharnuter W.M. 2003, A\&A 398, 1081

\end{thebibliography}
\end{document}